\documentclass{aa}

\usepackage[varg]{txfonts}

\usepackage{graphicx}
\usepackage{ifthen}
\usepackage{url}
\usepackage{amsmath}
\usepackage{bm}
\usepackage{lscape}
\usepackage{rotating}
\usepackage[graphicx]{realboxes}
\usepackage{supertabular}
\usepackage{pdfpages}
\usepackage{footnote}
\usepackage{hyperref} 
\usepackage{amssymb}
\usepackage{multirow}
\usepackage{xcolor}

\definecolor{myurlcolor}{HTML}{3B54A3}

\hypersetup{
  colorlinks   = true,    
  urlcolor     = myurlcolor,    
  linkcolor    = myurlcolor,    
  citecolor    = myurlcolor      
}

\newcommand{\lta}{$\; \buildrel < \over \sim \;$}
\newcommand{\simlt}{\lower.5ex\hbox{\ltsima}}
\newcommand{\gta}{$\; \buildrel > \over \sim \;$}
\newcommand{\simgt}{\lower.5ex\hbox{\gtsima}}

\def\kms{{\rm\,km\,s^{-1}}}

\def\masyr{{\rm\,mas\,yr^{-1}}}
\def\mmasyr{\,\mu{\rm as\,yr^{-1}}}
\def\kpc{{\rm\,kpc}}

\def\deg{^\circ}

\def\s{\ifmmode \widetilde \else \~\fi}
\def\={\overline}

\def\spose#1{\hbox to 0pt{#1\hss}}

\def\eg{{e.g.,\ }}
\def\ie{{i.e.,\ }}
\def\lta{\mathrel{\spose{\lower 3pt\hbox{$\mathchar"218$}}
     \raise 2.0pt\hbox{$\mathchar"13C$}}}
\def\gta{\mathrel{\spose{\lower 3pt\hbox{$\mathchar"218$}}
     \raise 2.0pt\hbox{$\mathchar"13E$}}}
\def\Dt{\spose{\raise 1.5ex\hbox{\hskip3pt$\mathchar"201$}}}    
\def\dt{\spose{\raise 1.0ex\hbox{\hskip2pt$\mathchar"201$}}}    

\def\dotsfill{\leaders\hbox to 1em{\hss.\hss}\hfill}

\begin{document}

\title{Reassessing the proper motions of M31/M33 with \textit{Gaia} DR3:}

\subtitle{Unraveling systematic uncertainties}

\author{Samuel Rusterucci \inst{1, 2}, Nicolas F. Martin \inst{1, 3}, Else Starkenburg \inst{2}, Rodrigo Ibata \inst{1}}

\institute{Université de Strasbourg, CNRS, Observatoire astronomique de Strasbourg, UMR 7550, F-67000 Strasbourg, France; \\ \email{samuel.rusterucci@astro.unistra.fr} \and Kapteyn Astronomical Institute, University of Groningen, Landleven 12, 9747 AD Groningen, The Netherlands \and Max-Planck-Institut für Astronomie, Königstuhl 17, D-69117 Heidelberg, Germany}
\date{Received 17 September 2024 / Accepted 21 October 2024}

\titlerunning{The proper motion of M31 $\&$ M33}
\authorrunning{Rusterucci et al.}

\abstract{
We provide an updated inference of the proper motion of M31 using the \textit{Gaia} DR3 proper motions of bright stars from the disc of M31. By refining the motion of the quasar reference frame, and statistically accounting for the variations in the inferred proper motions obtained across different regions of M31, we demonstrate that these inconsistencies most likely arise from systematic uncertainties. Our updated favoured values for the proper motion of M31 are $46.9\pm11.7(\mathrm{stat})\pm50.6(\mathrm{sys})\mmasyr$ along the right ascension direction, and $-29.1\pm9.4(\mathrm{stat})\pm35.6(\mathrm{sys})\mmasyr$ along the declination direction, the systematics being determined at a 90\% confidence level (the values for M33 are given in the paper). This clearly highlights that the systematics are the dominant source of uncertainty, their magnitudes being comparable to the proper motion of M31 itself. The analysis conducted using \textit{Gaia} DR2 instead of DR3 revealed that a net reduction in these systematic uncertainties occurred between the two data releases. If similar progress is made with the upcoming DR4, the future \textit{Gaia}-based estimates could match the level of uncertainties of HST, and could be used to refine the dynamics and history of M31 and M33.
}

\maketitle
\nolinenumbers

\defcitealias{2021MNRAS.507.2592S}{S21}

\section{Introduction}


The proper motion of the Andromeda galaxy (M31) is an essential piece of information to constrain the past, present, and future of the Local Group. While it is often assumed that most of the motion of M31 is carried by its radial velocity component \citep[\eg the timing argument;][]{1959ApJ...130..705K,2008MNRAS.384.1459L, 2016MNRAS.456L..54P}, useful measurements of its proper motion were only made possible over the last two decades from the use of indirect methods or significant advances in astrometric measurements. 

Indirect measurements of the proper motion of M31\footnote{Throughout this paper, all results are presented in the heliocentric reference frame to eliminate any dependency on the assumed motion of the Sun.} are primarily constrained by the radial velocities of its satellites, and were measured to be $(\mu_\alpha^\textrm{M31}, \mu_\delta^\textrm{M31}) = (21.5\pm11.1, -10.4\pm9.3)\mmasyr$ \citep{2008ApJ...678..187V} and $(9.1\pm19.0, 5.6\pm16.3)\mmasyr$ \citep{2016MNRAS.456.4432S}. These results, which are compatible with each other, when translated into the galactocentric frame of reference, display a substantial tangential motion that is not negligible compared to its radial counterpart. Nonetheless, these estimates rely on the assumption that, dynamically, M31 satellites are pressure supported and virialised, which may well not be the case given the observed presence of a plane of co-rotating satellites that includes about half of the known dwarf galaxies of M31 \citep{2013Natur.493...62I}. 

The direct measurements make use of the individual proper motions of stars within the disc of M31. \citet{2012ApJ...753....7S} and \citet{2012ApJ...753....8V} were the first to make such an attempt with highly accurate Hubble Space Telescope (HST) data from three deep fields ($3.37\arcmin \times 3.37\arcmin$), observed at two different epochs. Accounting for the internal kinematics of M31, they derived a weighted average of $(\mu_\alpha^\textrm{M31}, \mu_\delta^\textrm{M31}) = (44.1\pm12.7, -31.8\pm12.2)\mmasyr$ for the three fields, only marginally changing the results they obtained without considering the intrinsic motions of the stars within M31. This result is in much better agreement with a nearly radial orbit, but less so with the results from the indirect methods. They also derived a value by combining their direct measurement with indirect measurements to further constrain the motion of M31. Finally, the studies of \citet{2019ApJ...872...24V} and \citet[][hereafter S21]{2021MNRAS.507.2592S}, which respectively used the data from \textit{Gaia} DR2 and EDR3, concluded that the proper motion of M31 was also nearly radial in the latter study, but less so in the former. However, there are incompatibilities between the two samples analysed by \citetalias{2021MNRAS.507.2592S}. Their blue giant sample, which they consider more reliable due to its larger size and lower contamination, gives a nearly radial motion. In contrast, the red giant sample does not support this conclusion, even though it is inconceivable that two populations belonging to the disc of M31 move in different directions. Unfortunately, only hints could be offered in an attempt to explain these discrepancies, prompting us to revisit the study by \citetalias{2021MNRAS.507.2592S}, this time using data from \textit{Gaia} DR3, with a particular focus on refining the reference frame around M31.

\begin{figure*}
\centering
  \includegraphics[width=\textwidth]{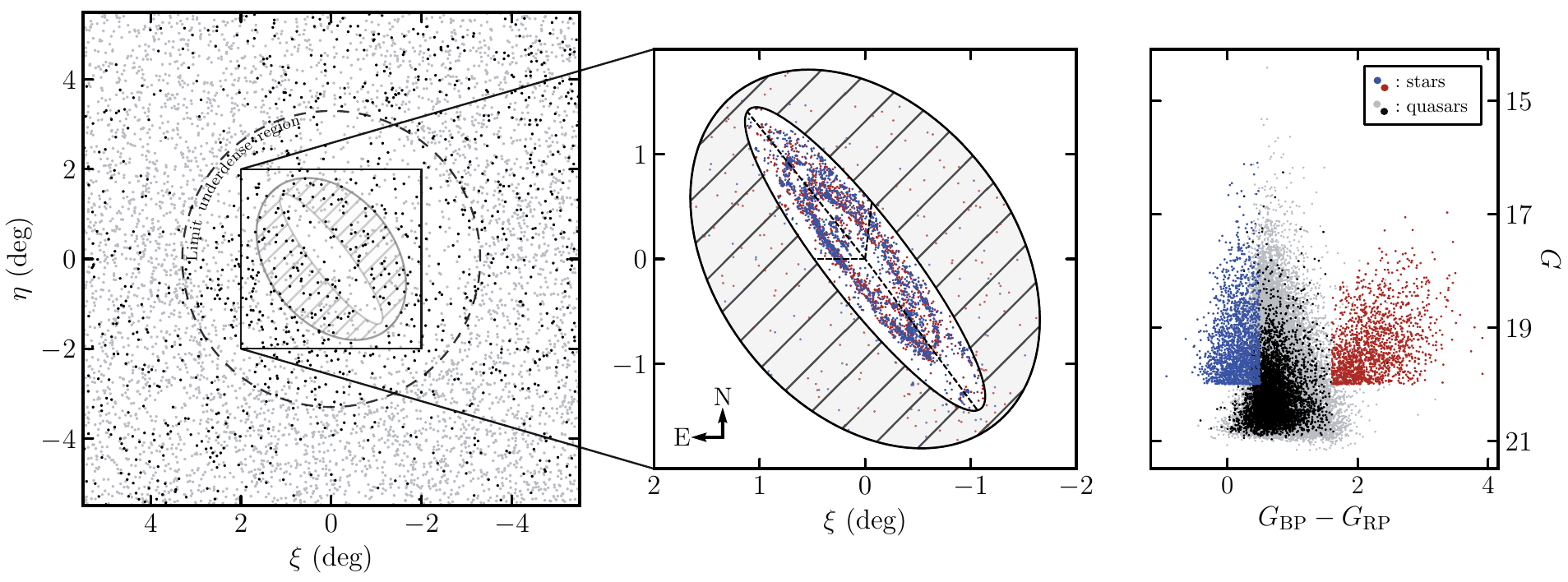}
  \caption{\textit{Left-hand panel:} quasars from the \textit{Gaia} CRF3 (in grey) and added by the latest \textit{Gaia} catalogue of extra-galactic objects (in black) centred on M31. \textit{Middle panel:} final blue and red samples of M31 stars superimposed onto the central ellipse representing the geometrical cut made on M31. Each quadrant depicted by the dashed lines contain the same number of blue stars and the hatched region is used to estimate properties of the contamination. \textit{Right-hand panel:} CMD of all the objects from the previous two panels.}
  \label{fig:stars_qsos}
\end{figure*}

Our motivation stems from the possibility of some systematic uncertainties on the \textit{Gaia} Celestial Reference Frame (CRF), that are unaccounted for at small scales. As it is shown in Figure 13 from \citet{2021A&A...649A...2L} and pointed out quantitatively in their Table 7, the CRF presents local uncertainties of the order of $\pm17.2\mmasyr$ for angular scales larger than 0.5$\deg$. At the distance of M31 this corresponds to $\pm63.9\kms$ \citep[assuming $d_{\textrm{M31}} = 785\pm25\kpc$,][]{2005MNRAS.356..979M}, which is on the order of magnitude of the previously measured tangential velocities. Here, we aim to refine the CRF by complementing the catalogue of quasars used to produce the \textit{Gaia} CRF, making use of the recently published extragalactic \textit{Gaia} catalogue \citep{2023A&A...674A..41G}. Given the very small values of the proper motion of M31, we also aim to carefully assess the level of systematic uncertainties present in the \textit{Gaia} data.

The structure of this paper is as follows: in Section~\ref{sec:method}, we describe our data and the statistical method used to infer the proper motion of M31. In Section~\ref{sec:result}, we derive an updated proper motion and estimate its systematic uncertainty. Finally, we discuss our results and conclude in Section~\ref{sec:discussion}.


\section{Methods}
\label{sec:method}
\subsection{Data selection}
\label{sub_sec:data_select}

For the full analysis, we use the \textit{Gaia} information provided in \citet{2023A&A...674A...1G}.

\textbf{Star samples:} To isolate M31 disc stars in the \textit{Gaia} catalogue we use the various spatial, colour-magnitude, proper motion\footnote{We follow the convention used in the \textit{Gaia} catalogue, noting $\mu_\alpha^*=\mu_\alpha\cos(\delta)$ as $\mu_\alpha$.} and \textit{Gaia} quality cuts (they are the same as in \citetalias{2021MNRAS.507.2592S}). The purpose of these cuts is to remove the contamination from foreground Milky Way (MW) stars and also stars with poor astrometric solutions. This leads to two separate samples of blue (young giant candidates; $B_{pm}$ sample) and red (older supergiant candidates; $R_{pm}$ sample) stars containing 1,867 and 1,543 objects, respectively. These stars are displayed on the sky and in the colour-magnitude diagram (CMD) in Figure~\ref{fig:stars_qsos}. Using the region surrounding the disc of M31 (hashed region in Figure~\ref{fig:stars_qsos}), we estimate that the two samples suffer from only low levels of contamination in agreement with the previous study (1.1\% and 1.8\%, respectively). The spatial distributions of the samples exhibit expected properties: stars from the blue sample are predominantly located in the ring of active star formation of the disc of M31 \citep{2015ApJ...805..183L} while the red stars are more sparsely distributed, but remain mainly confined to the region of the disc.

\textbf{Quasar sample:} Quasars used by the \textit{Gaia} consortium to produce the CRF of the DR3 are selected from a large sets of photometric and spectroscopic catalogues. These, however, may become heavily contaminated around nearby extended objects such as M31. In this case, only highly reliable quasars, confirmed through other means (\eg VLBI observations) were kept to constrain the reference frame \citep{2022A&A...667A.148G}. This leads to a drop in the density of quasars in the central $\sim3\deg$ region around M31 (grey dots in the left-hand panel of Figure~\ref{fig:stars_qsos}), which we can expect to locally lower the quality of the reference frame. To improve the latter, we complement the set of quasars used to build the CRF with additional objects selected from the latest \textit{Gaia} catalogue of extragalactic objects \citep{2023A&A...674A..41G}. In particular, we select sources that are highly likely to be quasars\footnote{Specifically, we select \textit{Gaia} quasar candidates flagged as \texttt{astrometric\_selection\_flag == 1}, \ie that have been identified with an estimated purity of 98$\%$ in \citet{2023A&A...674A..41G}.}, shown as black dots in Figure~\ref{fig:stars_qsos}. These newly added sources increase the central density from 19\,deg$^{-2}$ to 34\,deg$^{-2}$, compared to an average of 42\,deg$^{-2}$ over the whole sky\footnote{We also checked the quasar catalogue of \citet{2016ApJ...817...73H} based on the Pan-STARRS1 photometric variability of sources. Selecting likely quasars (\texttt{p\_QSO}$>$0.5) leads to no additional sources compared to quasars already added from the \textit{Gaia} extragalactic catalogue.}. Our final sample consists of 26,741 quasar candidates.

\begin{figure*}
\centering
  \includegraphics[width=\textwidth]{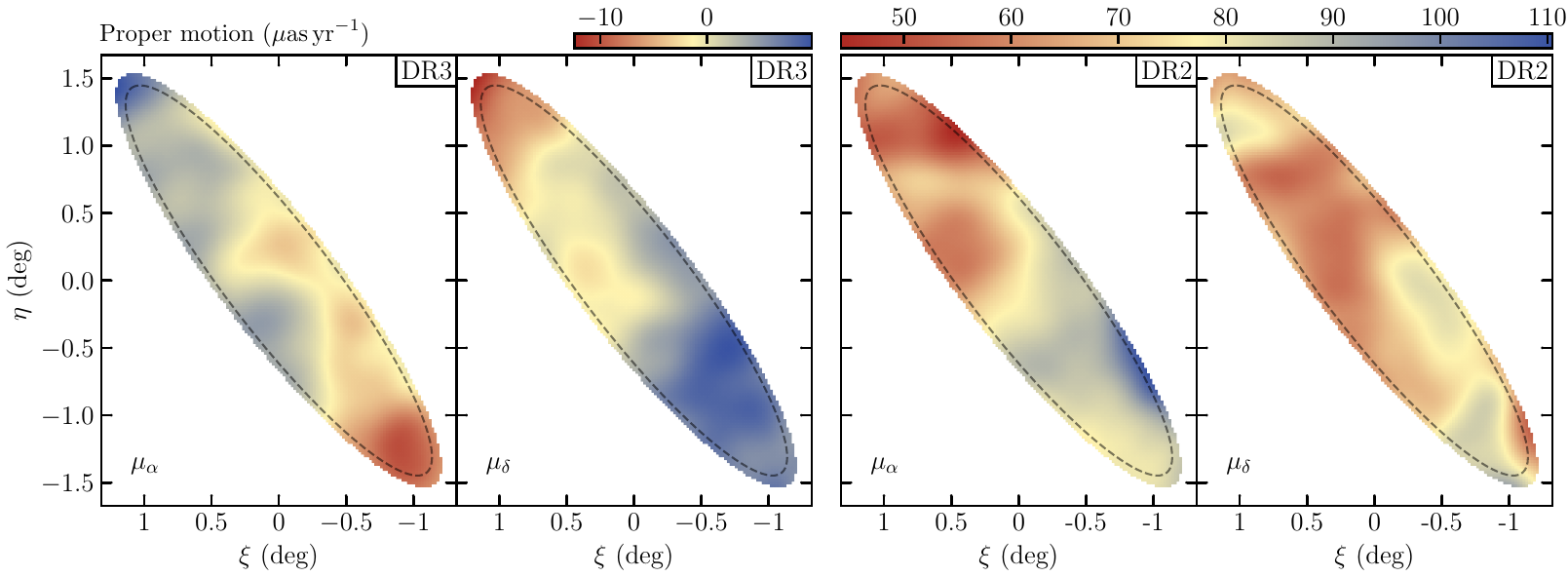}
  \caption{\textit{Left-hand panels:} smoothed maps of the quasar proper motions along the right ascension and the declination directions obtained using \textit{Gaia} DR3. These are used as zero-point offsets for the \textit{Gaia} stars proper motions. \textit{Right-hand panels:} same but using \textit{Gaia} DR2.}
  \label{fig:QSOs_Map}
\end{figure*}

\subsection{Proper motion inference}
\label{sec:prop_inf}

We aim to infer the global proper motion of stars in the \textit{Gaia} sample with a refined CRF. To do so, we assume that, for a given sample of objects (stars or quasars) of size $n$, we have a set of data points $\mathcal{D} = \{\mathbf{d}_{k}\}_{1\le k \le n}$, with datum $\mathbf{d}_k$ defined by the \textit{Gaia} position $(\alpha_k, \delta_k)$, proper motion $(\mu_{\alpha, k}, \mu_{\delta, k})$, along with their associated uncertainties\footnote{Given the vanishingly small uncertainties on the position of stars, we only consider the proper motion uncertainties.} $(\delta\mu_{\alpha, k}, \delta\mu_{\delta, k})$ and correlation coefficient $(\rho_k)$. That is: $\mathbf{d}_k = \{ \alpha_k, \delta_k, \mu_{\alpha, k}, \mu_{\delta, k}, \delta\mu_{\alpha, k}, \delta\mu_{\delta, k}, \rho_k \}$. The likelihood for all $n$ sources to follow a certain model specified by a set of parameters $\mathcal{P}$ is 

\begin{eqnarray}
P_\mathrm{tot}(\mathcal{D}|\mathcal{P})  = \prod_{k=1}^n P_k(\mathbf{d}_k|\mathcal{P}),
\end{eqnarray}    

\noindent where $P_\mathrm{tot}(\mathcal{D}|\mathcal{P})$ is the sum of two probabilistic models, $P_\mathrm{pop}$ and $P_\mathrm{cont}$, which represent the main population of M31 stars and contaminants, respectively.

\begin{eqnarray}
P_k(\mathbf{d}_k|\mathcal{P})  = (1-f_\mathrm{c})P_\mathrm{pop}(\mathbf{d}_k|\mathcal{P}^\mathrm{pop}) + f_\mathrm{c} P_\mathrm{cont}(\mathbf{d}_k|\mathcal{P}^\mathrm{cont})\\
\mathrm{with}\hspace{1.5mm} \mathcal{P}=f_\mathrm{c} \cup \mathcal{P}^\mathrm{pop} \cup \mathcal{P}^\mathrm{cont}, \hspace{3.62cm} \nonumber
\end{eqnarray}    

\noindent where parameter $f_\mathrm{c}$ represents the fraction of contaminants.

We choose to model $P_\mathrm{pop}$ as a two-dimensional Gaussian in the ($\mu_\alpha, \mu_\delta$) space, centred on the mean motion given by parameters $\mathcal{P}^\mathrm{pop} = \{\mu_\alpha^\mathrm{pop}, \mu_\delta^\mathrm{pop} \}$. The width of the Gaussian in the model is assumed to be entirely driven by the proper motion uncertainties and we forego parameters that would represent the intrinsic dispersion of sources\footnote{At the distance of M31, a dispersion of $100\kms$, would correspond to a contribution to the proper motion dispersion of only $\sim25\mmasyr$, which is much smaller than the typical proper motion uncertainties at the considered magnitudes.}. Therefore, we have

\begin{eqnarray}
        P_\mathrm{pop}(\mathbf{d}_k|\mathcal{P}^\mathrm{pop}) = \frac{1}{2\pi \delta\mu_{\alpha, k} \delta\mu_{\delta,k} \sqrt{(1 - \rho_k^2)}} \hspace{2cm}\nonumber\\ 
         \times \exp\bigg[ -\frac{1}{2(1-\rho_k^2)}\bigg(\frac{\Delta \mu_{\alpha,k}^2}{\delta\mu_{\alpha,k}^2} + \frac{\Delta \mu_{\delta,k}^2}{\delta\mu_{\delta,k}^2} - \frac{2\rho_k\Delta \mu_{\alpha,k}\Delta\mu_{\delta,k}}{\delta\mu_{\alpha,k}\delta\mu_{\delta,k}}\bigg)\bigg], 
    \label{eq:conditional_probability}
\end{eqnarray}

\noindent with $\Delta \mu_{\alpha,k}=\mu_{\alpha,k} - \mu_\alpha^\mathrm{pop}$ and $\Delta \mu_{\delta, k}=\mu_{\delta, k} - \mu_\delta^\mathrm{pop}$ being the offsets between a source's proper motion and the modelled motion of the considered population at this location. In the specific case of the stars of M31, a term accounting for the intrinsic motion of a star at position $k$ is subtracted to $\Delta \mu_{\alpha/\delta, k}$. The value of this term is determined from the motion predicted by the disc model developed by \citet{2009ApJ...705.1395C} for a star located at $(\alpha_k, \delta_k)$. In the case of the quasars, no additional term is needed as, on average, they are not expected to move.

We also choose $P_\mathrm{cont}$ to be modelled by a two-dimensional Gaussian. However, this time, the intrinsic dispersion of the contaminating stars cannot be assumed to be negligible anymore. The parameters are therefore the means and dispersions in the two-dimensional proper motion space, \ie $\mathcal{P}^\mathrm{cont} = \{ \mu_\alpha^\mathrm{cont}, \mu_\delta^\mathrm{cont}, \sigma_\alpha^\mathrm{cont}, \sigma_\delta^\mathrm{cont} \}$. Therefore,

\begin{eqnarray}
        P_\mathrm{cont}(\mathbf{d}_k | \mathcal{P}^\mathrm{cont})=\frac{1}{2\pi \sqrt{(\sigma_\alpha^\mathrm{cont})^2 + \delta\mu_{\alpha,k}^2}\sqrt{(\sigma_\delta^\mathrm{cont})^2 + \delta\mu_{\delta,k}^2}} \nonumber \hspace{-0.5cm} \\
	\times \exp\bigg[ -\frac{1}{2}\bigg(\frac{(\mu_{\alpha,k} - \mu_\alpha^\mathrm{cont})^2}{[(\sigma_\alpha^\mathrm{cont})^2 + \delta\mu_{\alpha,k}^2]} + \frac{(\mu_{\delta,k} - \mu_\delta^\mathrm{cont})^2}{[(\sigma_\delta^\mathrm{cont})^2 + \delta\mu_{\delta,k}^2]} \bigg)\bigg]. 
    \label{eq:conditional_probability}
\end{eqnarray}

Following Bayes' theorem, the posterior probability distribution function (PDF), $P_\mathrm{tot}(\mathcal{P}|\mathcal{D})$, is related to the likelihood $P_\mathrm{tot}(\mathcal{D}|\mathcal{P})$ through priors $P(\mathcal{P})$ such that

\begin{eqnarray}
P_\mathrm{tot}(\mathcal{P}|\mathcal{D})  \propto P_\mathrm{tot}(\mathcal{D}|\mathcal{P})P(\mathcal{P}).
\label{eq:likelihood}
\end{eqnarray} 

\begin{figure*}
\centering
  \includegraphics[width=\textwidth]{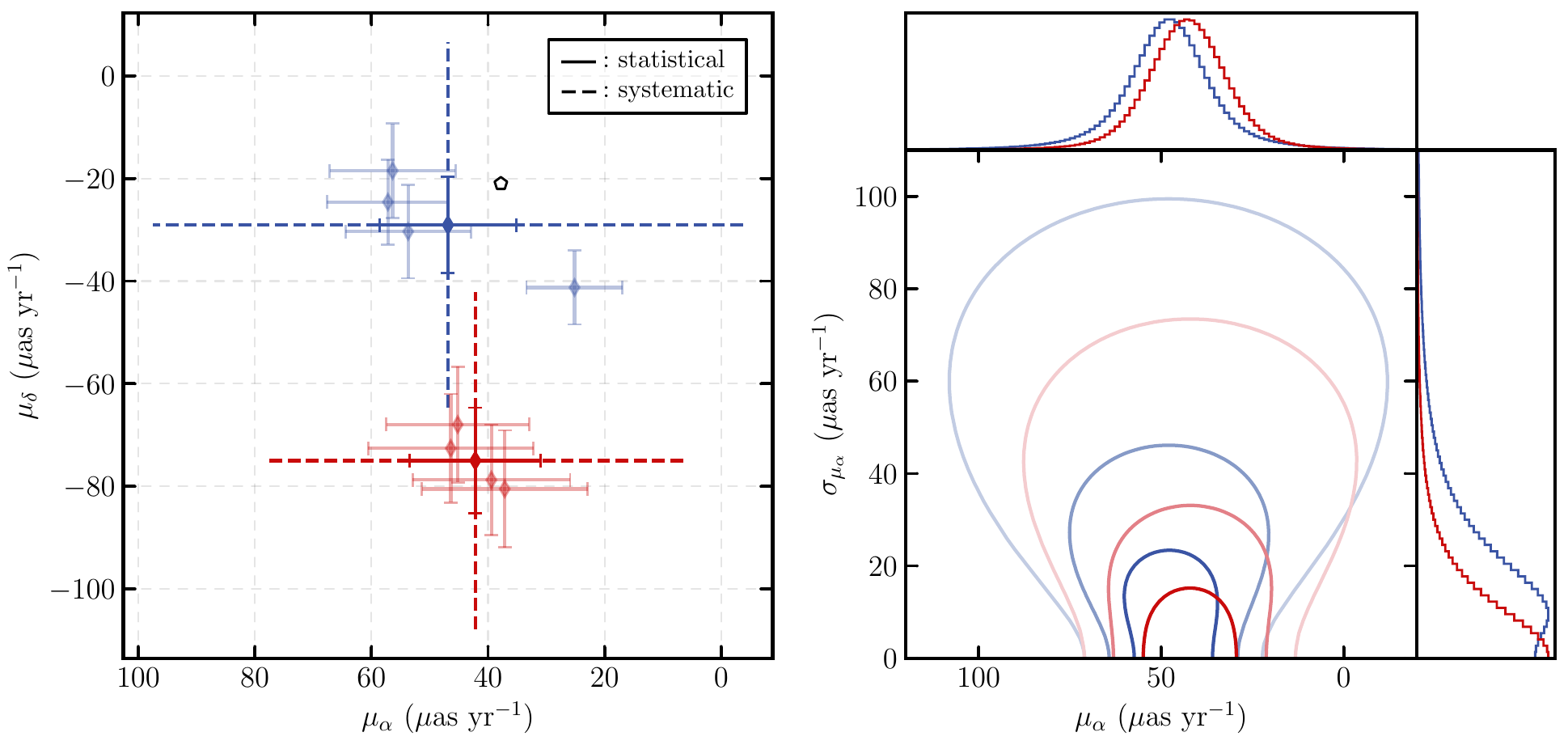}
  \caption{\textit{Left-hand panel:} results from the quadrant analysis of M31 in the heliocentric frame. In low-opacity are shown the inference obtained for each individual quadrant while in clear, are shown the inferences resulting from the analysis of the 2D space from the right-hand panel. The solide/dashed error bars respectively represent the statistical and systematic uncertainties. The black pentagon represents a strictly radial motion in the galactic reference frame. \textit{Right-hand panel:} example of the 2D PDFs (1, 2 and 3$\sigma$) obtained from the quadrant analysis along the right ascension direction and their marginalised 1D PDFs.}
  \label{fig:QuadAnalysis}
\end{figure*}

We assume uniform priors for all free parameters except for $f_\mathrm{c}, \sigma_\alpha^\mathrm{cont},$ and $\sigma_\delta^\mathrm{cont}$ for which a flat truncated prior is preferred. We choose $0\leq f_\mathrm{c}\leq 0.1$ as we estimated that the contamination for both samples of stars does not exceed 2$\%$. We choose $0\leq\sigma_\alpha^\mathrm{cont}, \sigma_\delta^\mathrm{cont}\leq2.5 \masyr$, as we estimated it to be $\sim1.5 \masyr$.

We sample the posterior PDFs using a Markov Chain Monte Carlo (MCMC) method with our own implementation of the Metropolis-Hastings algorithm \citep{1953JChPh..21.1087M, 1970Bimka..57...97H}.

\section{Results}
\label{sec:result}

\subsection{Correcting for the motion of the reference}
\label{sec:corr_ref}
The \textit{Gaia} data is of exceptional accuracy, however, at the distance of M31, motions of the order of a few tenth of $\mmasyr$ cannot be neglected. The motion of the \textit{Gaia} reference quasars is expected to be consistent with $0\mmasyr$. Nonetheless, there are reasons to expect that small systematics of the order of $10\mmasyr$ remain on small spatial scales \citep[see][Table 7]{2021A&A...649A...2L}. Here, we aim to quantify the level of these systematics so they can be folded into the analysis of the movement of M31 stars.

We first create a spatial grid with bins of sizes 72$''\times$72$''$, entirely covering M31 and parts of its outskirts. Using the method introduced in Section~\ref{sec:prop_inf}, we infer, for each bin, the mean motion of all quasars within a radius of $2.5\deg$ from the centre of the bin. This choice of radius serves two purposes: it makes sure that our approach remains as local as possible while also ensuring that each sample contains a minimum of 600 quasars. This number yields reasonable uncertainties of $\sim$10$\mmasyr$ on each inference. The resulting correction maps are then smoothed with a Gaussian kernel of standard deviation $\sigma = 5$\,pixels to reduce the noise caused by small-scale variations. We are aware that this strategy (using overlapping samples and then smoothing the maps) introduces correlations between the corrections in different bins, but, in turn, the resulting smoothed correction maps are simpler, more interpretable, and far less dominated by noise while maintaining high spatial resolution. The resulting quasar mean proper-motion maps are shown in the left-hand panels of Figure~\ref{fig:QSOs_Map} along the right ascension and the declination directions. While the maps show a mean motion close to $0\mmasyr$, they reveal peak to peak differences of up to $\sim20\mmasyr$. 

As expected from \citet{2021A&A...649A...2L}, the analysis using the DR3 proper motions shows significant improvement over the analysis using the DR2 proper motions. As an example, the right-hand panels of Figure~\ref{fig:QSOs_Map} show the corresponding analysis based on the DR2 data\footnote{We find only 23,139 DR2 counterparts to our DR3-based quasar sample, with the difference arising from the intrinsically more complete DR3 dataset \citep[see][]{2021A&A...649A...1G}.}. Not only is the range of corrections broader, reaching peak to peak differences of the order of $40\mmasyr$, but these corrections are systematically and significantly offset from $0\mmasyr$, with typical values of the order of the expected proper motion of M31 (see Table~\ref{tab:M31_Results}).

\begin{table*}[]
\caption{Proper motion of M31 in the heliocentric reference frame.}
\begin{center}
\setlength{\tabcolsep}{16pt}
\begin{tabular}{c l l c c}
\hline \hline
    \rule{0pt}{2ex} Sample & Study & Dataset & $\mu_{\alpha}^\mathrm{M31} (\mmasyr) $ & $\mu_{\delta}^\mathrm{M31} (\mmasyr) $\\ [\dimexpr 0.2ex] \hline
    & vdM19 & DR2  &   $65.0\pm24.0$  &  $-57.0\pm21.9 $  \\ \hline

                      & S21 & EDR3    &  $48.0\pm10.2$ & $-38.4\pm7.8$ \\  
$B_{pm}$    & This work & DR2      &  $-12.3\pm23.1(\mathrm{stat})\pm98.2(\mathrm{sys})$  &  $-37.7\pm27.1(\mathrm{stat})\pm122.3(\mathrm{sys})$ \\ 
                      & This work & DR3 &   $\mathbf{46.9\pm11.7(\mathrm{stat})\pm50.6(\mathrm{sys})}$  &  $\mathbf{-29.1\pm9.4(\mathrm{stat})\pm35.6(\mathrm{sys})}$  \\ \hline
                      
                      & S21 & EDR3    & $41.3\pm11.8$ & $-87.4\pm9.2$ \\
$R_{pm}$    & This work & DR2      &  $37.1\pm23.1(\mathrm{stat})\pm93.6(\mathrm{sys})$  &  $-134.8\pm20.7(\mathrm{stat})\pm79.1(\mathrm{sys})$    \\ 
                      & This work & DR3 &   $\mathbf{42.2\pm11.3(\mathrm{stat})\pm35.6(\mathrm{sys})}$  &  $\mathbf{-75.0\pm10.3(\mathrm{stat})\pm32.8(\mathrm{sys})}$
\end{tabular}
\tablefoot{vdM19 corresponds to \citet{2019ApJ...872...24V}. $B_{pm}$ and $R_{pm}$ are defined in S21 \citep{2021MNRAS.507.2592S} and in Section~\ref{sec:method} of this work.}
  \label{tab:M31_Results} 
\end{center}
\end{table*}

\begin{table*}[]
\caption{Proper motion of M33 in the heliocentric reference frame.}
\begin{center}
\setlength{\tabcolsep}{16pt}
\begin{tabular}{c l l c c}
\hline \hline
\rule{0pt}{2ex} Sample & Study & Dataset & $\mu_{\alpha}^\mathrm{M33} (\mmasyr) $ & $\mu_{\delta}^\mathrm{M33} (\mmasyr) $\\ [\dimexpr 0.2ex] \hline

        & vdM19 & DR2  &   $31.0\pm24.8$  &  $-29.0\pm22.6 $  \\ \hline
 \multirow{2}{*}{$B_{pm}^\textrm{M33}$}       & This work & DR2      &  $37.5\pm31.8(\mathrm{stat})\pm132.9(\mathrm{sys})$  &  $-47.7\pm28.4(\mathrm{stat})\pm121.6(\mathrm{sys})$ \\ 
                      & This work & DR3 &   $\mathbf{67.1\pm13.0(\mathrm{stat})\pm54.6(\mathrm{sys})}$  &  $\mathbf{-11.4\pm9.7(\mathrm{stat})\pm35.2(\mathrm{sys})}$  \\ \hline
                      
 \multirow{2}{*}{$R_{pm}^\textrm{M33}$}       & This work & DR2      &  $150.0\pm16.5(\mathrm{stat})\pm87.5(\mathrm{sys})$  &  $-46.6\pm26.1(\mathrm{stat})\pm102.3(\mathrm{sys})$ \\ 
                      & This work & DR3 &   $\mathbf{51.1\pm15.3(\mathrm{stat})\pm56.8(\mathrm{sys})}$  &  $\mathbf{11.4\pm11.9(\mathrm{stat})\pm50.0(\mathrm{sys})}$  \\ 
\end{tabular}
\vspace{2mm}
 \tablefoot{vdM19 corresponds to \citet{2019ApJ...872...24V}. $B_{pm}^\textrm{M33}$ and $R_{pm}^\textrm{M33}$ are defined in Section~\ref{sec:PM_M33} of this work.}
  \label{tab:M33_Results} 
\end{center}
\end{table*}

The DR3 correction maps can finally be applied as zero-point offsets to the \textit{Gaia} proper motions of stars in the $B_{pm}$ and $R_{pm}$ samples. This approach is more local than the correction applied by \citetalias{2021MNRAS.507.2592S}, who determined a mean offset over a region of 4 to 20$\deg$ around M31. Applying the method described in Section~\ref{sec:prop_inf} yields $(\mu_\alpha^\mathrm{M31},\mu_\delta^\mathrm{M31})=(44.6\pm 14.9, -29.8\pm 13.1)\mmasyr$ for $B_{pm}$ and $(41.3\pm 14.3, -74.8\pm 13.2)\mmasyr$ for $R_{pm}$. The uncertainties\footnote{All statistical uncertainties throughout this paper are determined as the central 68-percent confidence interval.} are here computed as the quadratic sum of the uncertainty resulting from the inference itself and the mean uncertainty from the inferences on the quasar corrections ($9.8\mmasyr$ along the right ascension direction and $8.3\mmasyr$ along the declination direction). These results are similar to those obtained by \citetalias{2021MNRAS.507.2592S} for the same samples, including the stark difference along the declination direction between the blue and red samples. Despite our more local correction of the mean motion of quasars, this measurement is still plagued by systematics that we now aim to constrain.

\subsection{Quadrant analysis}
\label{sec:quad_analysis}
To explore the reliability of the mean proper motion of M31 inferred above, we divide our blue and red samples into four quadrants centred on M31 (delimited by the dashed lines in the central panel of Figure~\ref{fig:stars_qsos}), each containing the same number of stars, and then infer the bulk motion of these smaller but independent samples. Consistent results in the different quadrants would mean that the systematics are folded into the statistical uncertainties whereas incompatible results between quadrants would imply additional systematics that are not yet taken into account.

The results for each quadrant are shown as the low-opacity symbols and associated error bars in Figure \ref{fig:QuadAnalysis}. For this particular quadrant configuration, the results for the blue stars are inconsistent, one quadrant significantly offset from the three others. As we rotate the quadrants by small steps, we see rapid changes in the individual proper motions, leading to this inconsistency that is not specifically driven by our choice of quadrants. Potential contamination contributing to these inconsistencies is discussed in Section \ref{sec:contamination}.


To assess the level of potential systematics in the proper motion inferences, we assume that the results from the four quadrants correspond to four independent measurements of the proper motion of M31\footnote{As already mentioned above in Section \ref{sec:corr_ref}, this is not entirely accurate due to the correlations introduced by our method of constructing the quasar correction maps and subsequently smoothed them.}, whose PDF on the mean ($\mu_{\alpha/\delta}$) and dispersion ($\sigma_{\mu_{\alpha/\delta}}$) informs us on the mean proper motion of the considered sample of stars and the unaccounted systematics. An example of such PDFs are shown in right-hand panel of Figure~\ref{fig:QuadAnalysis} for the right ascension proper motion, with the corresponding colours for the blue and red sample of stars.

\begin{figure*}
\centering
  \includegraphics[width=\textwidth]{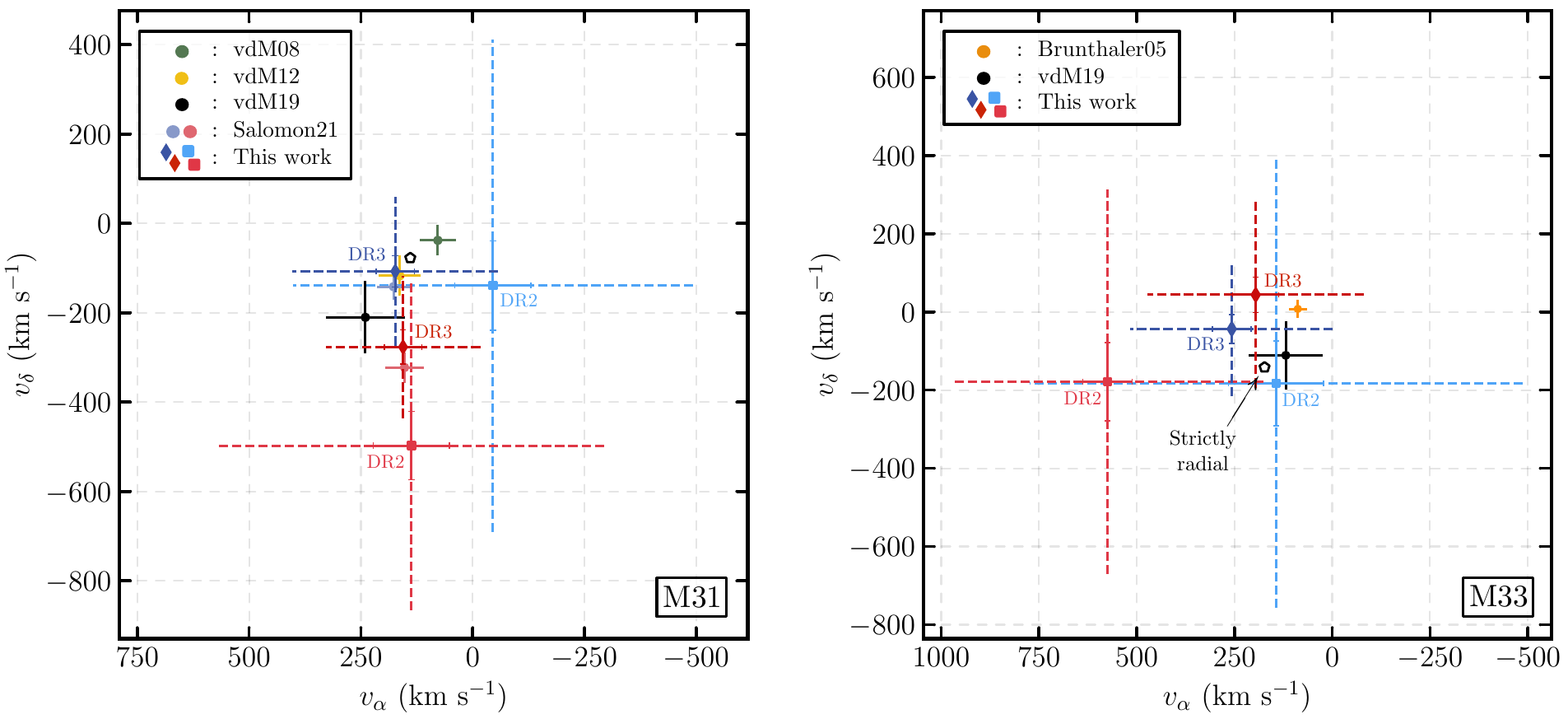}
  \caption{\textit{Left-hand panel:} proper motions of M31 in the heliocentric reference frame. The black pentagon represents a strictly radial motion in the galactic reference frame. The yellow dot is the weighted average of the three HST fields corrected for the internal kinematics of M31 \citep{2012ApJ...753....7S, 2012ApJ...753....8V}. The black dot is the proper motion result based on the DR2 \citep{2019ApJ...872...24V}. The blue and red dots are the values derived by \citet{2021MNRAS.507.2592S} while the diamonds and squares in different shades of blue/red are the results from this paper using the DR3 and the DR2, respectively. For our study, the solid error bars represent the statistical uncertainty and the dashed ones the systematic uncertainties. \textit{Right-hand panel:} proper motions of M33 in the heliocentric reference frame. The orange dot is the water maser weighted average of \cite{2005Sci...307.1440B}. The rest of the symbols are the same as on the left-hand panel.}
  \label{fig:FinalResults}
\end{figure*}

An estimate of the mean and its statistical uncertainty are obtained from the marginalised PDFs of ($\mu_{\alpha/\delta}$), while the systematics are determined from the marginalised PDFs of ($\sigma_{\mu_{\alpha/\delta}}$). Conservatively, we choose to determine the systematics on the proper motion as the 90-percent confidence limit on this parameter. The corresponding values are shown as the dashed error bars in the left-hand panel of Figure~\ref{fig:QuadAnalysis}. The final results, detailing the statistical/systematic uncertainties, are listed in Table~\ref{tab:M31_Results} and displayed in the left-hand panel of Figure~\ref{fig:FinalResults}. In both cases they are compared to literature values. The offset mentioned throughout this paper, between the blue and red stars, remains in our analysis. However, the systematics dominate the overall uncertainties, rendering the two star samples compatible with one another, which was not the case before (the potential origins of those will be further discussed in Section~\ref{sec:discussion}). Moreover, these results are also compatible with the previous estimate of \citet{2019ApJ...872...24V} who used \textit{Gaia} DR2.

\subsection{The proper motion of M33}
\label{sec:PM_M33}
M33 is a companion of M31, located roughly at the same distance \citep[$d_{\textrm{M33}} = 794\pm24\kpc$, from][]{2004MNRAS.350..243M}. It also shows signs of active star formation \citep{2007MNRAS.379.1302B, 2024MNRAS.52710668P}, which, as in the case of M31, implies the presence of super-giants bright enough to be detected by \textit{Gaia}. Before \textit{Gaia}, estimates of the proper motion of M33 had already been made thanks to the presence of two water masers in its disc, IC 133 and M33/19, whose positions could be very accurately followed through interferometric VLBI measurements \citep{2005Sci...307.1440B}. Although we can be highly confident in the precision of these measurements, much like the HST measurements of M31 \citep[performed in small deep fields,][]{2012ApJ...753....7S}, this estimate may be model-dependent and affected by local peculiar motions. As for the case of M31, an independent measurement, albeit more uncertain, was made possible through \textit{Gaia} DR2 \citep{2019ApJ...872...24V}. The results from these previous works are shown in the right-hand panel of Figure \ref{fig:FinalResults} as the orange and black dots, respectively.

We isolate M33 stars by applying the same cuts as the ones used for M31. The resulting blue ($B_{pm}^\textrm{M33}$) and red ($R_{pm}^\textrm{M33}$) samples are made of 1,992 and 1,093 stars, respectively. We correct for the intrinsic motion of the stars with the flat disc model of \citet{1997ApJ...479..244C} and then follow the whole procedure described in the previous sections. That is, obtaining the quasar correction maps and use them as zero-point offsets to the \textit{Gaia} proper motions. These maps are very similar to the ones of M31, having mean motions compatible with 0$\mmasyr$ (within the uncertainties of each bin inference) and peak to peak differences of $\sim10\mmasyr$. 

The quadrant analysis reveals inconsistencies between the quadrant inferences, similar to those observed for M31, and leading to similarly large systematic uncertainties. The final results are displayed in the right-hand panel of Figure~\ref{fig:FinalResults} and are listed in Table~\ref{tab:M33_Results}. Interestingly, contrary to the case of M31, there is a better agreement between the $B_{pm}^\mathrm{M33}$ and $R_{pm}^\mathrm{M33}$ samples which are nearly compatible within their statistical uncertainties. This might be the result of a less contaminated CRF around M33 than around M31, or that the analysis is conducted on a smaller area (M33 being smaller). Our results also appear mostly consistent with the previous works, displaying an overall proper motion slightly less radial towards the MW than the result found by \citet{2019ApJ...872...24V}.


\section{Discussion}
\label{sec:discussion}
\subsection{Summary}
We reproduce the results from the study that was made by \citetalias{2021MNRAS.507.2592S}. In order to correct for the background motion of the quasars more locally, we created correction maps which revealed local differences of $\sim$10$\mmasyr$ \citep[as expected from][]{2021A&A...649A...2L}. Applying these correction maps as zero-point offsets for the stars yields results very similar to those previously found, and despite the corrections, there remains an observed offset between the inferences of the blue and the red samples of stars. This offset is most likely due to systematics present in the \textit{Gaia} DR3 data. 

We show that we are able to assess the level of those systematics from the analysis of the overall inferred motion of different regions of the disc of M31. With a conservative 90-percent confidence level, these systematics are of the order of $30\mmasyr$ and commensurate with the proper motion of M31 itself. Such large systematics, explained by the spread of results from M31 star sub-samples, help to decipher the observed difference between the blue and the red sample of stars, that was first pointed out by \citetalias{2021MNRAS.507.2592S}. These new results render entirely compatible our new estimates with the ones made by \citet{2019ApJ...872...24V} (using \textit{Gaia} DR2) and \citet{2012ApJ...753....7S}  and \citet{2012ApJ...753....8V} (using HST), these all being displayed in the left-hand panel of Figure~\ref{fig:FinalResults}. No proper motion estimate of M33 had been made using \textit{Gaia} DR3, but our updated results are consistent with previous works on the topic (see right-hand panel of Figure~\ref{fig:FinalResults}). 

Overall, it is, of course, unsatisfactory to be unable to tightly constrain these systematics that are large compared to the proper motions of either M31 or M33, but nevertheless, very small and close to the accuracy expectations of \textit{Gaia}. Below, we discuss potential sources that could explain these unaccounted systematics. 

\subsection{Impact of colour}
The strong difference in the inferred proper motion of M31 from the blue and red samples of stars appear to imply that colour could (in part) play a role in the presence of systematics. \citet{2021A&A...649A...4L}, in their study of the dependence in colour and magnitude of the parallax systematics in \textit{Gaia} DR3, pointed out that, since parallaxes and proper motions are jointly determined, such (small) systematics are likely also present for the \textit{Gaia} proper motion values. \citet{2021A&A...649A.124C} showed that, at the bright end, these were indeed biased, with the biases being a function of the magnitude, the position and the colour of an object. They estimated that the systematics could be of the order of $10\mmasyr$, for objects that are bluer or redder than the mean colour of \textit{Gaia} objects. The colour-magnitude diagram in the right-hand panel of Figure \ref{fig:stars_qsos} shows the colour distribution of stars from both samples along with that of the background quasars. It is clear that there is no strong overlap between the different populations, which may explain why, if there is indeed a colour-dependent systematic to the \textit{Gaia} proper motions over the full magnitude range, the quasar correction did not remove this effect.


\subsection{Potential contamination}
\label{sec:contamination}
With the accuracy that we are aiming for, biases may arise from differences between the assumed model and the true distribution of the data. In particular, it is easy to imagine that the presence of a single outlier with small proper motion uncertainties (\eg a bright contaminating star) may shift the mean inferred proper motion. This shift may be small overall, but might still be comparable to the tiny mean proper motion of M31.

As already pointed out by \citetalias{2021MNRAS.507.2592S}, the PAndAS MW stream, located at a distance of $\sim 17 \kpc$ \citep{2014ApJ...787...19M} and crossing M31 in its most northern part, overlaps with the blue sample in the CMD for $G<18$. However, re-doing the analysis only for stars with $G>18$ does not change our conclusions.

We also investigate the intrinsically higher contamination of the red sample that was already mentioned in \citetalias{2021MNRAS.507.2592S}. Using the numerous photometric bands of the PHAT survey \citep{2014ApJS..215....9W} and, in particular, the $(F336W-F475W)$ vs. $(F110W - F160W)$ colour-colour space\footnote{The first combination is gravity-sensitive while the latter explores the infrared colours of the stars.}, we see a clear separation between stars removed by the proper motion cut and the $R_{pm}$ sample. We interpret this as the separation between contaminating dwarfs and relevant giants. This separation, if truly discriminating the contaminants from true members would imply that the 1.8\% contamination (from Section~\ref{sub_sec:data_select}) would be underestimated at least by a factor of two, the newly found contamination fraction being of the order of 4\% for the $R_{pm}$ sample. Re-doing the full analysis without these possible contaminants highlighted in the PHAT data barely changes our previous results, the new inference remaining far from the blue inferences. 


\subsection{Improvements since \textit{Gaia} DR2 and future prospects}
\label{sec:DR2}

Because aiming at proper motion precisions challenging \textit{Gaia}'s current capabilities, it is worth checking whether the situation has improved between the DR2 and the DR3. To this end, when available, we extract the DR2 proper motions for the quasar, the $B_{pm}$ and $R_{pm}$ samples built in Section~\ref{sub_sec:data_select}. We then perform the full analysis on this new DR2-based dataset. The right-hand panels of Figure \ref{fig:QSOs_Map} show the corresponding quasar correction maps, which as already mentioned in Section~\ref{sec:corr_ref}, is significantly worse than for the DR3: not only are the corrections no longer centred on $\sim0\mmasyr$, but the full amplitude of corrections spans $\sim 40\mmasyr$ (vs. $\sim0\mmasyr$ and $\sim20\mmasyr$, respectively, for the DR3). The quadrant inconsistencies are larger than for the DR3 analysis, implying larger final systematic uncertainties reaching $\sim90\mmasyr$, two to three times larger than for the DR3 analysis (see Table \ref{tab:M31_Results}). In Figure~\ref{fig:FinalResults}, we see that these results have statistical uncertainties comparable to the ones from \citet{2019ApJ...872...24V}. However, we see that the model's favoured values are very different and are only compatible because of the large systematic uncertainties. We observe the same trend for M33 where the results using the DR2 were way more uncertain than the ones using the DR3.

As noted by \citet{2021A&A...649A...2L}, the proper motion uncertainties should decrease over time as $T^{-3/2}$, that is by factor of 0.51 between the DR2 (22 months of data) and the DR3 (34 months of data). It is reassuring to observe this refinement in our results, both for the statistical and systematic uncertainties. If a similar improvement is achieved between the DR3 and the DR4, the proper motion uncertainties should be reduced by nearly a factor of 3. We hope this will help resolve the discrepancies between the blue and red star samples. We also anticipate the possibility of reaching systematic uncertainties comparable to the $\sim12\mmasyr$ reached by HST \citep[in][]{2012ApJ...753....7S, 2012ApJ...753....8V}, but over a sample of stars covering the entire extent of M31. Re-evaluating the proper motions of M31 and M33 with the DR4 could further constrain their dynamics, allowing for studies of their orbital history to be compared with thorough past works \citep[\eg][]{2012ApJ...753....9V, 2017MNRAS.464.3825P}.

\begin{acknowledgements}
      This work has made use of data from the European Space Agency (ESA) mission {\it Gaia} (\url{https://www.cosmos.esa.int/gaia}), processed by the {\it Gaia} Data Processing and Analysis Consortium (DPAC, \url{https://www.cosmos.esa.int/web/gaia/dpac/consortium}). Funding for the DPAC has been provided by national institutions, in particular the institutions participating in the {\it Gaia} Multilateral Agreement. 
\end{acknowledgements}

\bibliographystyle{aa}
\bibliography{refs}

\end{document}